\documentclass[twocolumn,showkeys,superscriptaddress,preprintnumbers,amsmath,amssymb,showpacs,prl]{revtex4-1}
\usepackage{graphicx}% Include figure files
\usepackage{dcolumn}% Align table columns on decimal point
\usepackage{bm}% bold math
\usepackage{amsmath}
\usepackage{xcolor}

\begin{document}

\title{Seismicity in sheared granular matter}

\author{Aghil Abed Zadeh}
\affiliation{Department of Physics \& Center for Non-linear and Complex Systems, Duke University, Durham, NC, USA}
\author{Jonathan Bar{\'e}s}
\affiliation{Laboratoire de M\'{e}canique et G\'{e}nie Civil, Universit\'{e} de Montpellier, CNRS, Montpellier, France}
\author{Joshua E.~S.~Socolar}
\affiliation{Department of Physics \& Center for Non-linear and Complex Systems, Duke University, Durham, NC, USA}
\author{Robert P.~Behringer}
\affiliation{Department of Physics \& Center for Non-linear and Complex Systems, Duke University, Durham, NC, USA}

% In this Letter, we experimentally investigate granular avalanches and their statistics beyond conventional power-laws. We use time declustering to analyze the sequence of avalanches and study avalanches correlation using certain conditional probabilities. We find  crackling dynamics for low driving rate, evolving into a more predictable sequence of avalanches for higher driving rates. This Letter may address a wide range of researches on granular physics, geophysics, material plasticity, and crackling noise in many other systems. Moreover, our results can provide direct comparison with seismic avalanches data to understand the role of granular materials in those.

%%%#$#%%%#$#%%%#$#%%%#$#%%%#$#%%%#$#%%%#$#%%%#$#%%%#$#%%%#$#%%%#$#%%%#$#%%%#$#%%%#$#%%%#$#%%%#$#%%%#$#%%%#$#%%%#$#%%%#$#%%%#$#%%%#$#%%%

\begin{abstract}
We report on experiments investigating the dynamics of a slider that is pulled by a spring across a granular medium consisting of a vertical layer of photo-elastic disks. The motion proceeds through a sequence of discrete events, analogous to seismic shocks, in which elastic energy stored in the spring is rapidly released.  We measure the statistics of several properties of the individual events: the energy loss in the spring, the duration of the movement, and the temporal profile of the slider motion. We also study certain conditional probabilities and the statistics of mainshock-aftershock sequences. At low driving rates, we observe crackling with Omori-Utsu,  B\r{a}th, and waiting time laws similar to those observed in seismic dynamics. At higher driving rates, where the sequence of events shows strong periodicity, we observe scaling laws and asymmetrical event shapes that are clearly distinguishable from those in the crackling regime.
\end{abstract}

\date{\today}



\maketitle

%%%#$#%%%#$#%%%#$#%%%#$#%%%#$#%%%#$#%%%#$#%%%#$#%%%#$#%%%#$#%%%#$#%%%#$#%%%#$#%%%#$#%%%#$#%%%#$#%%%#$#%%%#$#%%%#$#%%%#$#%%%#$#%%%#$#%%%

A wide variety of physical systems exhibit stick-slip dynamics in response to a steady driving force. In such systems, configurations of microscopic elements remain stable as the force builds up, then undergo fast microscopic rearrangements during macroscopic relaxation events. When the driving force is increased slowly, the distribution of event sizes often develops a power-law form, and the system is said to ``crackle'' \cite{Sethna2001_nat}.  Examples can be found in phenomena associated with fracture \cite{Bonamy2008_prl,Bares2014_prl,Ribeiro2015_prl,Bares2018_natcom}, friction \cite{Brace1966_sci,Johnson2008_nat}, magnetization \cite{Urbach1995_prl}, neural activity \cite{Beggs2003_jn,Bellay2015_elife} and seismicity \cite{Bak2002_prl,Davidsen2013_prl,Bares2018_natcom}, to mention a few. 

Many experimental studies \cite{Dalton2001_pre,Baldassarri2006_prl,Petri2008_epjb,Denisov2017_sr,Bares2017_pre,Murphy2018_arx}, simulations \cite{Maloney2004_prl,Salerno2012_prl,Otsuki2014_pre,Liu2016_prl,Bares2017_pre}, and theoretical models \cite{Dahmen2011_nat,Talamali2011_pre,Budrikis2013_pre,Lin2014_pnas} have investigated the statistics of sizes and durations slip events in disordered athermal systems. Such studies typically focus on the quasi-static regime, where the driving rate is slower than any relevant intrinsic time scale. Several conceptual tools for characterizing the dynamics were originally developed for the purpose of analyzing seismological data and forecasting earthquakes. In that context, it has been observed that the statistics of mainshocks ($MS$) and aftershocks ($AS$) follow several phenomenological scaling laws \cite{Arcangelis2016_pr}: the Gutenberg-Richter power-law decay in the event size distribution; a power-law distribution of the waiting times between successive events \cite{Bak2002_prl,Castellanos2018_arx,Bares2018_natcom}; a power-law distribution of $AS$ productivity as a function of $MS$ magnitude \cite{Utsu1971_jfs,Helmstetter2003_prl}; B\r{a}th's law \cite{Bath1965_tec} for the constancy of the size ratio of the largest $AS$ to the $MS$ that produced it; and the Omori-Utsu law \cite{Omori1894_jcsiut,Utsu1972_jfs,Utsu1995_jpe} for the decay in the rate of $AS$ following a given $MS$. These laws, referred to as the fundamental laws of seismology \cite{Ogata1988_jasa}, are used in the probabilistic forecasting of earthquakes, and it is important to understand their mechanistic origins and potential applicability to other types of crackling phenomena. 

This paper reports on experimental studies of a slider that is pulled over a granular material. The system undergoes a transition from crackling to periodic dynamics as the driving rate is increased~\cite{Zadeh2018_arx}, and we present results for all of statistical quantities mentioned above in both regimes. We find that for slow driving rates our system follows familiar seismic laws and their associated predictions. For higher driving rates, however, the statistics of plastic events changes significantly. Our results in the crackling regime can serve as tests for general theories that have been developed in other contexts, such as the theory of epidemic-type aftershock sequences~\cite{Ogata1988_jasa,Helmstetter2003_grl}.

%%%#$#%%%#$#%%%#$#%%%#$#%%%#$#%%%#$#%%%#$#%%%#$#%%%#$#%%%#$#%%%#$#%%%#$#%%%#$#%%%#$#%%%#$#%%%#$#%%%#$#%%%#$#%%%#$#%%%#$#%%%#$#%%%#$#%%%

\begin{figure}
\centering \includegraphics[width=0.95\columnwidth]{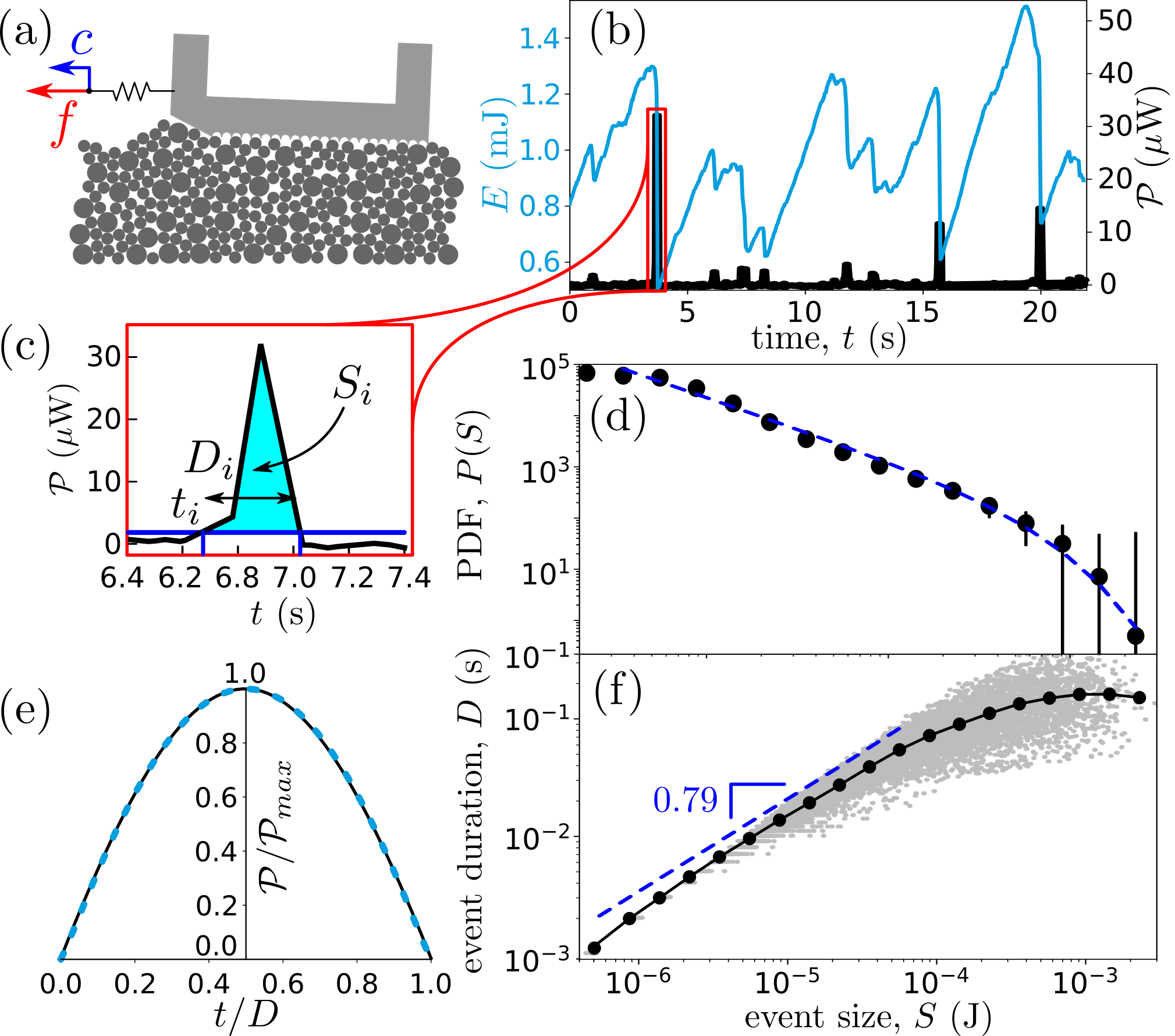}
\caption{(color online) 
%Event data for loading speed $c=0.1$~mm/s. (a): Time series of the energy $E$ stored in the loading spring (blue) and the radiated power $\mathcal{P}$ (black). Peaks in $\mathcal{P}$ represent slip events. \jb{Inset: zoom on a peak and schematic of the threshold based method used to extract avalanches (see text).} (b): PDF of the event size $S$ as measured by the energy drop in the spring. The blue dashed line is a fit to the form of Eq.~(\ref{eqRG}) with $\beta=1.22 \pm 0.07$ and $S_{max}=6.1 \times 10^{-4} \pm 0.7 \times 10^{-4}\,$J. Error bars show $95$\% confidence intervals. Inset: Schematic of the experimental set-up (see text). (c): Event duration, $D$, as a function of event size, $S$. Gray dots are individual events; the black line is the binned average. The dashed straight line shows a power-law with exponent $1/\gamma= 0.79 \pm 0.03$. Inset: Average event shape for $D \in [0.01,0.02]$~s. The black solid line is computed from the experimental data, and the blue dashed line is a fit of eq.\ref{eqSHP} with $\sigma = 1.09 \pm 0.02$.
Stick- slip event statistics for loading speed $c=0.1$~mm/s. (a) Schematic of the experiment at loading speed $c$. (b) Time series of the spring energy $E$ (blue) and radiated power $\mathcal{P}$ (black). 
%Peaks in $\mathcal{P}$ represent slip events. 
(c) An enlarged view of a single peak in $\mathcal{P}$ with starting time $t_i$, duration $D_i$, and size $S_i$. The horizontal line indicates the threshold used for event detection. (d) PDF of the event sizes.
%$S$ as measured by the energy drop in the spring. 
The blue dashed line is a fit to the form of Eq.~(\ref{eqRG}) with $\beta=1.22 \pm 0.07$ and $S_{max}=6.1 \times 10^{-4} \pm 0.7 \times 10^{-4}\,$J. Error bars show $95$\% confidence intervals. (e) Average event shape for $D \in [0.01,0.02]$~s. The black solid line is computed from the experimental data, and the blue dashed line is a fit to Eq.~(\ref{eqSHP}) with $\sigma = 1.09 \pm 0.02$. (f) Event duration as a function of event size. Gray dots are individual events; the black line is the binned average. The dashed line shows a power-law corresponding to $1/\gamma= 0.79$.}
\label{fig1}
\end{figure}

{\it Experimental procedure:\/}
The experimental set-up used for the present study is similar to the one used in Ref.~\cite{Zadeh2018_arx,Zadeh2017_epj}, illustrated schematically in Fig.~\ref{fig1}(a). A toothed 2D frictional slider of length $25$~cm and mass $85$~g is attached to a linear spring of stiffness $k=70$~N/m. The end of the spring is pulled at a constant speed $c$. The slider lies on a vertical bed of bi-disperse cylindrical elastic particles of diameters $4$~mm and $5$~mm, with the number ratio $N_{\text{small}}/N_{\text{big}}=2.7$. The granular bed and the slider are sandwiched between two dry-lubricated glass plates. The force, $f$, applied to the spring is measured by a sensor at a frequency of $1$~kHz. From the time series $f(t)$, we calculate the elastic energy stored in the spring, $E=f^2/(2k)$, and the power radiated during a given time step, $\mathcal{P}$. In our experiments, $c$ can vary from $0.02$ to $100$~mm/s.

Slip events, also called plastic events or avalanches, are identified directly from the $\mathcal{P}(t)$ data. To avoid including transient effects, we consider the force signal only after the slider has moved approximately half of its length.  After this point the signal is seen to oscillate around a well defined average. For slower pulling speeds, we collect roughly 1,000 avalanches per run, and we include several runs in our statistical analysis.  For higher speeds, we collect ten runs with roughly 50 slip events per run.

To identify avalanches, we choose a threshold for $\mathcal{P}$ of $1.5\,\mu$W, which is above our rms noise level of $1.16\,\mu$W, as illustrated in Fig.~\ref{fig1}(c), an avalanche starts when $\mathcal{P}$ exceeds the threshold and ends when it next drops below the threshold~\cite{Bares2014_fp,Bares2017_pre}. For each event we extract the starting time, $t_i$, the duration, $D_i$ and the event size $S_i$, defined as the total energy radiated during the event; and the temporal profile of the power released by the spring, $\mathcal{P}_i(\tau)$ for $\tau \in [0,D_i]$, where $\tau = t - t_i$.
We then construct the following statistical quantities: the probability density function (PDF) for event sizes, $P(S)$; the average duration of events of a given size, $D(S)$; and the average scaled event shape for events with $D_i \in [D-\delta,D+\delta]$,  $\overline{\mathcal{P}_D}(u) \equiv \langle \mathcal{P}_i(\tau)/ \max(\mathcal{P}_i(\tau)) \rangle_i$, where $u=\tau/D_i$.
Finally, we identify the $AS$'s associated with a given $MS$ for the purpose of observing B\r{a}th's law, the productivity law, and the Omori-Utsu law.

%%%#$#%%%#$#%%%#$#%%%#$#%%%#$#%%%#$#%%%#$#%%%#$#%%%#$#%%%#$#%%%#$#%%%#$#%%%#$#%%%#$#%%%#$#%%%#$#%%%#$#%%%#$#%%%#$#%%%#$#%%%#$#%%%#$#%%%

\begin{figure}
\centering \includegraphics[width=0.7\columnwidth]{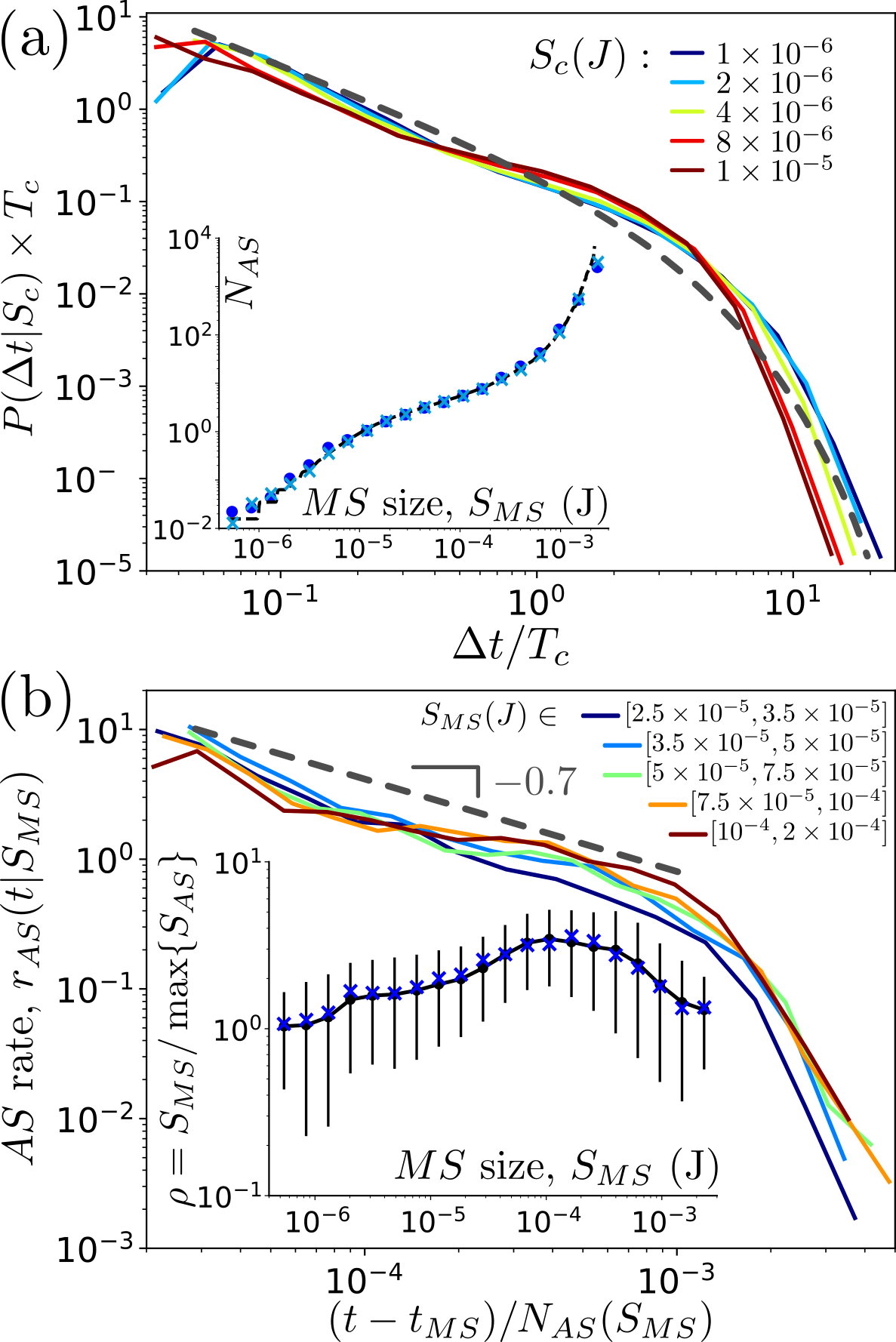}
\caption{(color online) Waiting time law, Omori-Utsu law, and B\r{a}th's for $c=0.1$~mm/s. (a) Waiting time law: PDF of the waiting time between two consecutive events larger than $S_{c}$, rescaled by $T_{c}$, the average waiting time between events larger than $S_{c}$. The dashed line is a fit to the form of Eq.~(\ref{eqWT}) with $\nu= 1.10 \pm 0.08 $ and $u_{max}= 3.04 \pm 0.15$. Inset: The number of $AS$'s as a function of the $MS$ energy, $S_{MS}$. Blue dots are from experiment; cyan crosses are obtained from random permutations of the experimental events, and the black dashed line is obtained from Eq.~(\ref{eqPRD}). (b) The rate of $AS$ events corresponding to a $MS$ of energy $S_{MS}$. Time is rescaled by $N_{AS}(S_{MS})$. The straight dashed line with slope $0.7$ is a guide to the eye. Inset: Ratio between $MS$ size, $S_{MS}$, and size of the largest $AS$. Black dots are experimental data; cyan crosses are obtained from a randomly mixed sequence of events.
}
\label{fig2}
\end{figure}

{\it Crackling dynamics:\/}
When the slider is pulled very slowly, we observe crackling dynamics, as evidenced for $c=0.1\,$mm/s by the two decades of power-law decay in $P(S)$ shown in Fig.~\ref{fig1}(d). The PDF is well fit by the Gutenberg-Richter form
\begin{equation}
	P(S) \sim S^{-\beta} e^{-\frac{S}{S_{max}}},
\label{eqRG}
\end{equation}
with $\beta=1.22 \pm 0.07$ and $S_{max}=(6.1 \pm 0.7) \times 10^{-4}\,$~J. Fig.\ref{fig2}(c) shows a scatter plot of the size and duration of individual events. The average curve $\langle D(S)\rangle$ also exhibits a power-law, $\langle D\rangle \sim S^{1/\gamma}$, over two decades, with $1/\gamma= 0.79 \pm 0.03$. For large events, the average duration appears to saturate.  

Fig.~\ref{fig1}(e) shows the average shape of avalanches with $D_i \in [0.01,0.02]\,$s, which places them cleanly in the power law regime. Each avalanche has been rescaled vertically by its maximum power radiated and horizontally by its duration. The form of $\overline{\mathcal{P}_D}(u)$ has been investigated for a variety of amorphous systems \cite{Dahmen2011_nat,Liu2016_prl,Bares2017_pre,Denisov2017_sr}. We follow the procedure of Refs.~\cite{Laurson13_natcom} and~\cite{Bares2014_prl}, fitting the average shape to the form:
\begin{equation}
	\overline{\mathcal{P}_D}(u)=4[u(1-u)]^{\sigma}[1-a(u-1/2)]
\label{eqSHP}
\end{equation}
where $a$ is an asymmetry parameter. We find $a=0$, and $\sigma = 1.09 \pm 0.02$. We note that $\sigma$ differs significantly from the value $\sigma=\gamma-1$ predicted in Ref.~\cite{Laurson13_natcom}.

%%%#$#%%%#$#%%%#$#%%%#$#%%%#$#%%%#$#%%%#$#%%%#$#%%%#$#%%%#$#%%%#$#%%%#$#%%%#$#%%%#$#%%#%%%#$#%%%#$#%%%#$#%%%

{\it Correlations between events:\/}
We now consider the statistical relations between avalanches. Fig.~\ref{fig2}(a) shows the waiting time distribution, which would show a simple exponential decay for an uncorrelated Poisson process. Each curve is obtained by setting a lower cutoff size $S_{\rm c}$ and compiling the PDF of the times $\Delta t = t_{j'} - t_j$, where $j$ and $j'$ refer to successive events with $S > S_{\rm c}$.  The curves are scaled horizontally by the average waiting time, $T_{\rm c}$, between the included events, which yields a collapse to a single curve that can be fit by a power law with an exponential cutoff \cite{Bak2002_prl}:
\begin{equation}
	P(\Delta t | S_{c}) \sim \frac{1}{T_{c}}u^{-\nu} e^{-\frac{u}{u_{max}}},
\label{eqWT}
\end{equation}
where $u=\Delta t/T_{c}$. We find $\nu= 1.10 \pm 0.08$ and $u_{max}= 3.04 \pm 0.15$.  The dashed line shows the best fit. %\az{The power-law dependence of this law evidences temporal correlation between events, as it cannot be described by an uncorrelated Poisson process with exponential distribution of waiting time.}

Another conventional method for characterizing correlations between events is known as seismicity declustering~\cite{Baro2013_prl,Bares2018_natcom}.  For any event $M$ (called the main shock), associated aftershocks are defined as all subsequent events occurring before the next event larger than $M$. The inset of Fig.~\ref{fig2}(a) shows the average number, $N_{AS}$ of $AS$ associated with $MS$ of a given size, $S_{MS}$. We have checked that his productivity law remains unchanged after random shuffling of the sequence of events. Given the absence of relevant correlations in the sequence of event sizes, the law can be computed directly from the cumulative event size distributions $F(S)=\int_{S_{min}}^S P(s)ds$ \cite{Bares2018_natcom}:
\begin{equation}
	N_{\rm AS}(S_{\rm MS}) = \frac{F(S_{\rm MS})}{1-F(S_{\rm MS})}\,.
\label{eqPRD}
\end{equation}
The result, shown as a dashed line in the inset of Fig.~\ref{fig2}(a), matches the direct computation, confirming that the sequence of event sizes is essentially random.

Finally, we study the temporal structure of aftershocks, which is characterized by the Omori-Utsu law for the time dependence of the rate $r$ of aftershock events \cite{Omori1894_jcsiut,Utsu1995_jpe} and B\r{a}th's law for the ratio $\rho$ of the size of a $MS$ to its largest $AS$. Fig.~\ref{fig2}(b) shows the rate $r$ of aftershocks following a $MS$ of size $S_{MS}$ as a function of the time elapsed since the $MS$, averaged over $MS$ with similar sizes: $r$ is a function of $(t-t_{MS})$. Curves for different $S_{MS}$ collapse under the rescaling $t-t_{MS} \rightarrow (t-t_{MS})/N_{AS}(S_{MS})$ revealing that:
\begin{equation}
	r(t | S_{MS}) \sim u^{-p}f(u)
\label{eqOMO}
\end{equation}
where $u=(t-t_{MS})/N_{AS}(S_{MS})$ and $f$ is a scaling function. We find $p \approx 0.7$.  

The inset of Fig.~\ref{fig2}(b) shows $\rho$ as a function of $S_{MS}$. Within the error bars, the data is consistent with B\r{a}th's law, which states that $\rho=S_{MS}/\max\{S_{AS}\}$ is independent of $S_{MS}$. Once again, permuting the sequence of events randomly does not change the curve, as shown by blue crosses. This implies that, like productivity law, B\r{a}th's law finds its origin in the distribution of individual event sizes, without requiring system memory of the type introduced in epidemic-type aftershock sequence (ETAS) models \cite{Ogata1988_jasa,Helmstetter2003_grl}.

%%%#$#%%%#$#%%%#$#%%%#$#%%%#$#%%%#$#%%%#$#%%%#$#%%%#$#%%%#$#%%%#$#%%%#$#%%%#$#%%%#$#%%%#$#%%%#$#%%%#$#%%%#$#%%%#$#%%%#$#%%%#$#%%%#$#%%%

\begin{figure}
\centering \includegraphics[width=0.7\columnwidth]{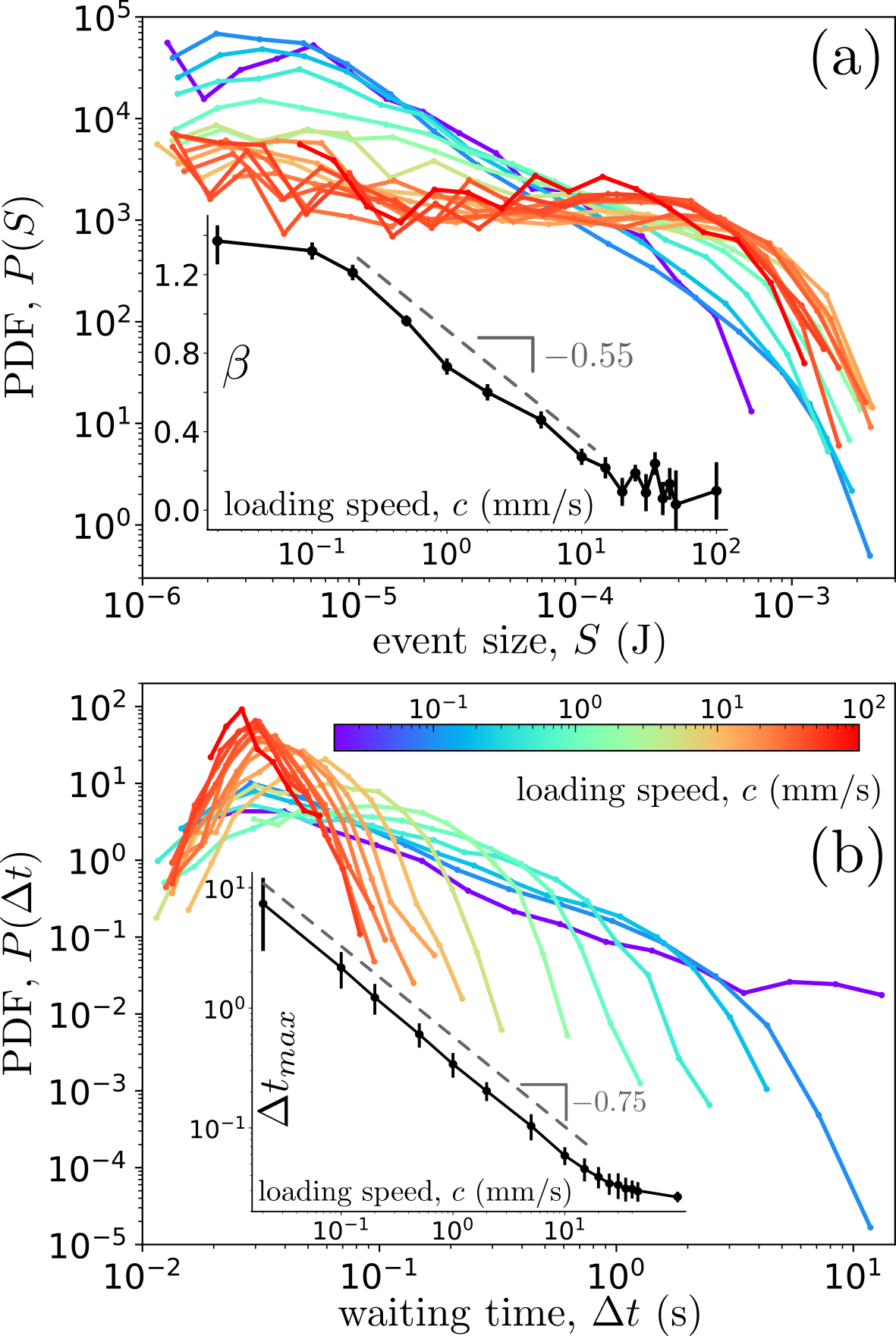}
\caption{(color online) (a): PDF of the event energy, $P(S)$, for loading speeds $c \in [0.02,100]$~mm/s. Inset: Logarithmic decay of the exponent $\beta$ with increasing $c$. The fit for $10^{-1}<c<20$~mm/s has slope $-0.55 \pm 0.06$. (b): PDF of the waiting time between two consecutive events larger than $S_{\rm th}=10^{-6}$~J for different $c$. Inset: Evolution of the upper cut-off, $\Delta t_{max}$, (see Eq.~(\ref{eqWT})) as a function of $c$. The fitted power-law for $c<20$~mm/s shows an exponent of $-0.75 \pm 0.03$.}
\label{fig3}
\end{figure}

%\section{Dependence on driving rate}
{\it Dependence on driving rate:\/}
As the driving rate is increased, the system passes into the periodic regime discussed in Ref.~\cite{Zadeh2018_arx}. We report here the changes in avalanche statistics associated with this transition. Figure~\ref{fig3}(a) shows the evolution of $P(S)$ with increasing $c$. The crackling behavior at low driving rates evolves to a uniform distribution (with an upper cut-off) reaching a stable form for $c>c_c \approx 20\,$mm/s. The inset shows the dependence on $c$ of the $P(S)$ exponent, $\beta$, indicating that $\beta \sim \log(c)$  for $10^{-1}<c<c_c$. For faster $c$, $\beta$ vanishes within the error bars. For slower $c$ it saturates at a constant value near $1.22$.

Figure~\ref{fig3}(b) shows the evolution of $P(\Delta t)$ to a narrow, Gaussian-like distribution for $c>c_c$ indicating a characteristic time between events. The inset shows the upper cut-off, $\Delta t_{max}$, as a function of $c$, revealing a power-law decay for $c$ slower than $c_c$: $\Delta t_{max} \sim c^{-0.75 \pm 0.03}$. For faster $c$, $\Delta t_{max}$ decays only logarithmically with increasing $c$, as discussed in Ref.~\cite{Zadeh2018_arx}.

\begin{figure}
\centering \includegraphics[width=0.7\columnwidth]{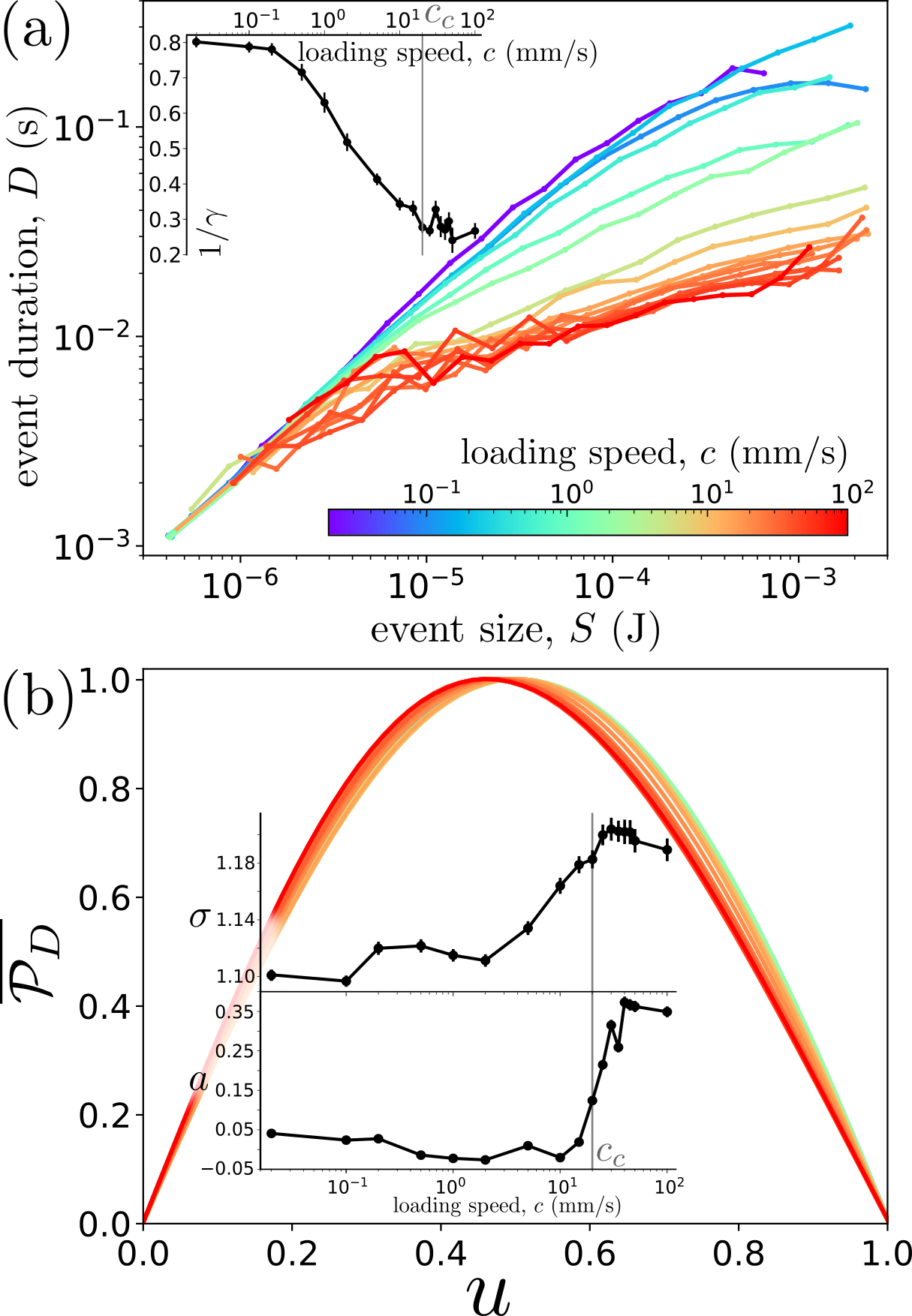}
\caption{(color online) (a): Evolution of avalanche duration, $D(S)$, for loading speeds $c \in [0.02,100]$~mm/s. Inset: Evolution of the exponent, $1/\gamma$. (b): Average shape $\overline{\mathcal{P}}_D$ for events with $D_i \in [0.01,0.02]$s. Colors correspond to the scale in (a). Insets: Evolution of the exponent, $\sigma$, and asymmetry parameter, $a$, for event shapes as a function of $c$ showing that events develop left-shifted asymmetry as $c$ increases. In all graphs error bars stand for $95$\% confidence level and vertical gray lines show $c=c_c=20$~mm/s.}
\label{fig4}
\end{figure}

Both $D(S)$ and $\overline{\mathcal{P}_D}(u)$ for $D = 0.015\,$s show clearly discernible differences in the periodic regime from their crackling behavior at low $c$. Figure~\ref{fig4}(a) shows $D(S)$ for different pulling speeds. While $D(S)$ is a power-law in all cases, the exponent is observed to decrease from $\sim 0.8$ for low $c$ to $\sim 0.3$ for $c>c_c$.  

Figure~\ref{fig4}(b) shows how the average event shape of Fig.~\ref{fig1}(e) evolves with increasing $c$. $\overline{\mathcal{P}_D}(u)$ develops a left-shifted asymmetry as $c$ increases. The exponent $\sigma$ and asymmetry parameter $a$ of Eq.~(\ref{eqSHP}) are plotted as a function of $c$ in the top and bottom insets. $\sigma$ is roughly constant for low $c$, then slowly increases to a saturation value for $c>c_c$. $a$ is near zero for low $c$ and increases sharply to a constant for $c>c_c$, meaning that during the average event in the periodic regime, the slider accelerates faster than it decelerates. 

%%%#$#%%%#$#%%%#$#%%%#$#%%%#$#%%%#$#%%%#$#%%%#$#%%%#$#%%%#$#%%%#$#%%%#$
%\section{Discussion}
{\it Discussion:\/}
Our experiments demonstrate that slowly shearing a granular bed by means of a pulled slider on its surface gives rise to the same types of scale free distributions of event sizes, durations, and shapes as those seen in other crackling systems. We observe behavior that is well described by standard versions of the Gutenberg-Richter law, Omori's law, the productivity law, and B\r{a}th's law. As the driving rate is increased, the parameters characterizing the statistics vary smoothly until a crossover occurs to the qualitatively different behavior observed in the periodic regime reported in Ref.~\cite{Zadeh2018_arx}.

Comparing our results to those from other systems in the crackling regime, we find the following. Our size exponent $\beta$ matches that observed in many other studies.   Several quasi-static theoretical models and simulations \cite{Talamali2011_pre,Salerno2012_prl,Budrikis2013_pre,Lin2014_pnas,Liu2016_prl}, as well as fracture experiments by Bar{\'e}s et al.~\cite{Bares2017_pre}, reproduce the same scaling exponent within the error bars. This value is smaller, however, than that observed in other dynamic 3D experiments \cite{Dalton2001_pre,Baldassarri2006_prl,Lherminier2016_rcf,Denisov2017_sr} and the mean field model exponent of $1.5$ \cite{Dahmen2011_nat,Otsuki2014_pre,Denisov2017_sr,Castellanos2018_arx}, and it is larger than was found some other studies \cite{Maloney2004_prl,Ispanovity2014_prl}. The exponent $\gamma$ we measure for the $D(S)$ power-law agrees with 3D experiments of Dalton et al.~\cite{Dalton2001_pre} and with dislocation dynamics models \cite{Ispanovity2014_prl}. It is lower, however, than those of simulations by Liu et al.~\cite{Liu2016_prl} and a mean field model by Budrikis et al.~\cite{Budrikis2013_pre}. Our measured exponent $\nu$ for the waiting time law agrees with the creep mean field model \cite{Castellanos2018_arx}. Finally, as has been recently reported for nominally brittle fracture in amorphous materials \cite{Bares2018_natcom}, we find that the productivity statistics and B\r{a}th's law can be directly implied from the Gutenberg-Richter law, without knowing the correct sequence of events. This fact disqualifies ETAS~\cite{Ogata1988_jasa} for modeling our system.

The symmetric parabolic avalanche shape we observe in the crackling regime is predicted by many models \cite{Laurson13_natcom}, including mean field \cite{Dahmen2011_nat}. In the periodic regime, however, we find asymmetric avalanche shapes with parameters different from those predicted by 3D simulations \cite{Liu2016_prl}. The source of the asymmetry is not clear, but we conjecture that it is related to a crossover in time scales associated with the slider motion and the relaxation dynamics of the granular bed. Moreover, we measure an exponent $\sigma$ different from the value $\gamma-1$  predicted in Ref.~\cite{Laurson13_natcom}. This may be due to the combination of the physical effect of a finite driving rate and the experimental resolution that requires a finite threshold value for event detection.  The former may yield overlapping avalanches that we classify as a single event, and the latter may result in the splitting of single avalanches into two or more events~\cite{Janicevic2016_prl}. Similar features were observed in Barkhausen pulses \cite{Papanikolaou2011_nat} and were shown to result from the finite value of the demagnetization factor. 
Further investigation of the grain scale dynamics during slip events is needed to clarify the origin of these macroscopic phenomena.

%%%#$#%%%#$#%%%#$#%%%#$#%%%#$#%%%#$#%%%#$#%%%#$#%%%#$#%%%#$#%%%#$#%%%#$#%%%#$#%%%#$#%%%#$#%%%#$#%%%#$#%%%#$#%%%#$#%%%#$#%%%#$#%%%#$#%%%

{\it Acknowledgements:\/} We gratefully acknowledge fruitful discussions with Kirsten Martens and Daniel Bonamy.  We also acknowledge the crucial role played by Bob Behringer in supporting and mentoring this research. Though he died unexpectedly before this manuscript could be drafted, his contribution clearly justifies inclusion as a coauthor. The work was supported by NSF grants DMR-1206351 and DMR-1809762, NASA grant NNX15AD38G, the William M. Keck Foundation, and DARPA grant 4-34728.

%%%#$#%%%#$#%%%#$#%%%#$#%%%#$#%%%#$#%%%#$#%%%#$#%%%#$#%%%#$#%%%#$#%%%#$#%%%#$#%%%#$#%%%#$#%%%#$#%%%#$#%%%#$#%%%#$#%%%#$#%%%#$#%%%#$#%%%

\bibliographystyle{unsrt}
\bibliography{biblio.bib}

\end{document}